# The Age Distribution of Potential Intelligent Life in the Milky Way

Daniel Legassick

Student ID: 610003136

University of Exeter

## Abstract

We investigated the habitability of the Milky Way, making use of recent observational analysis on the prevalence of Earth-sized planets, in order to estimate where and when potentially habitable star systems may have formed over the course of the Galaxy's history. We were then able to estimate the age distribution of potential intelligent life in our Galaxy using our own evolution and the age of the Sun as a proxy. To do this we created a galactic chemical evolution model and applied the following habitability constraints to the Sun-like (G-type) stars formed in our model: an environment free from life-extinguishing supernovae, a high enough metallicity for Earth-sized planet formation and sufficient time for the evolution of complex life. We determined a galactic habitable zone as the region containing all the potentially habitable star systems in our model. Our galactic habitable zone contains stars formed between 11 and 3.8 billion years ago at radial distances of between 7 and 14 kiloparsecs. We found that most potentially habitable star systems are much older than the Sun and located farther from the galactic centre. By comparing the ages of these systems we estimated that ~77% of potentially habitable star systems are on average ~3.13 billion years older than the Sun. This suggests that any intelligent life in the Galaxy is likely to be incredibly more advanced than we are assuming that they have evolved under similar timescales than we have. Implications and limitations of our study are discussed.

# **Table of Contents**

# § 1 Introduction

Are we alone in the Universe? This is one of the most important questions facing our civilisation today and has no doubt been pondered by countless people over the centuries. The sheer size of the universe leads many to conclude that, "no, we are not alone". However, with no obvious evidence of other civilisations' existence, this reality seems difficult to accept. Therefore, at present this question should be approached from a scientific point of view – looking at what we know about how life came about here on Earth and applying it to our entire Galaxy to see how common the conditions for life are elsewhere. In this study we have attempted to do just that, by creating a detailed Galaxy simulation which we have used to investigate where and when intelligent life could have potentially emerged.

### § 1.1 The Galactic Habitable Zone

Central to this research area is the concept of a galactic habitable zone (GHZ), which was first proposed by Gonzalez, et al. (2001). The GHZ is defined as the area of the Galaxy which may favour the emergence of complex life (Gowanlock et al. 2011) and has been predicted to be an annular region which widens with time (Lineweaver et al. 2004). However, Prantzos (2008) found that almost all of the Galaxy's disk may contain habitable planets at present, therefore the necessity for a GHZ is less motivated. The main factors which are thought to determine the location of the GHZ are: Supernovae, Metallicity, Planet formation and the timescale for biological evolution. The Inner boundary of the GHZ is thought to be defined by the larger stellar density at smaller radial distances from the galactic centre, meaning an increased frequency of high energy events such as supernovae, which can destroy planetary biospheres (Gonzalez et al. 2001, Lineweaver et al. 2004). Whereas the outer boundary is thought to be defined by the minimum metallicity required for planet formation (Gonzalez et al. 2001, Lineweaver et al. 2004).

### § 1.2 The Present Study

The main purpose of the previous GHZ studies has been to explore where habitable planets may have formed with enough time for complex life to develop in the Galaxy (where complex life is defined as land-based animal life (Gowanlock et al. 2011)). The main purpose of this study on the other hand is to estimate how old intelligent civilisations which may have evolved on some of these planets may be in comparison to us. Thereby estimating the age distribution of potential intelligent life in the Galaxy. Our motivation for conducting this study was the fact that this question had not been explored in significant detail previously, which is surprising considering that the data required to conduct a study such as this has existed for some time. However, in recent years analysis of exoplanetary data has yielded invaluable estimates of the prevalence of Earth-sized planets in our Galaxy (Buchhave et al. 2012, Petigura et al. 2013), which now allows us to better constrain the occurrence of such systems in galactic habitability studies such as this.

Keeping with the GHZ studies we have modelled Supernovae, Metallicity and the timescale for intelligent life to evolve. We have not modelled planet formation as recent research has very accurately calculated the fraction of Sun-like stars which harbour Earth-sized planets in the habitable zone (HZ) of their host star, based on de-biased observational data of exoplanets (Petigura et al. 2013). By applying this fraction to the stars in our model we negate the need to model planet formation or indeed the inclusion of planets at all - saving us valuable computing time. This approach is justified as the time difference between the formation of a star and the formation of its planets is negligibly small in comparison to the lifetime of the star.

# § 2 Methodology

In order to estimate how the number of potentially habitable planets has evolved over different epochs of the Galaxy's history we created a galactic chemical evolution model (GCEM) using the computer programming language MATLAB (MathWorks, 2015) to determine which areas of the Galaxy's disk may have been the most hospitable to life and when. This GCEM incorporated interstellar gas, star formation, supernovae and metallicity. The radial component of our simulated Galaxy covered the range 2.5-15 kiloparsecs (kpc), whilst time varied from the formation of the Galaxy (~13.6 billion years ago (Pasquini et al. 2004)) to present day. Therefore, our simulation yielded a two-dimensional model which allowed us to study habitability as a function of time and radial distance from the galactic centre.

We did not model the very centre of the Galaxy at R < 2.5 kpc, as the bulge overlaps the disk here (Gowanlock et al. 2011) and the evolution of the bulge is not as well constrained as the rest of the disk (Prantzos, 2008). We only modelled up to 15 kpc for consistency with Gowanlock et al. (2011) and because beyond this radius the density of the disk stars appears to drop rapidly to zero (Sparke and Gallagher, 2007). Additionally, we did not model the vertical component of the Galaxy, Z, due to limited computing power and because the majority of potentially habitable planets are thought to be located within the Galaxy's disk which has a relatively small vertical distribution.

In the following sections I will explain how we set up our GCEM, how we used it to predict where and when habitable star systems may have formed and how we determined the age distribution of potential intelligent civilisations in the Galaxy from this.

## § 2.1 Distribution of Gas

The first thing we needed to do in our GCEM was to set up the initial distribution of gas at the Galaxy's formation, t=0. The gas surface density decreases exponentially with radial distance from the centre of the Galaxy (Naab and Ostriker, 2006).

$$\Sigma_d(r,t) = \Sigma_0(t) e^{-r/r_d(t)} \qquad (1)$$

Here $r_d$ is the scale length: the distance you have to move radially from the centre of the Galaxy until the gas surface density decreases by a factor e, and $\Sigma_0$ is the central surface density of the Galaxy. The scale length is estimated to be between 2500-3500 parsecs (pc) (Naab and Ostriker, 2006), 2500-4500 pc (Sparke and Gallagher, 2007) and 2000-3000 pc (Binney and Tremaine, 2008) at the present day. However, it is difficult to measure because of our location in the disk (Binney and Tremaine, 2008) and the scale length at the formation of the Galaxy is unknown. Therefore, $r_d$ as well as $\Sigma_0$ are included as variables in our model which are chosen to best match the present day observed quantities of the Galaxy. This equation is used to assign different masses of gas to each radial bin at the first timestep, t=0 and is the only initial condition in our model.

During a galaxies lifetime it converts gas into stars. The rate at which stars form from the interstellar gas is called the star formation rate (SFR). In the next section I will discuss the SFR we chose to use in our GCEM.

**§ 2.2 Star Formation Rate**

A popular method for modelling star formation is using the Schmidt or Schmidt-Kennicutt law. This is a simple relationship between the surface densities of gas and star formation, first introduced by Schmidt (1959).

$$\sum \text{SFR} = A \sum gas^N$$

From this equation we can see that the gas mass lost as it is turned into stars (per parsec squared per year) is proportional to the total mass of gas (per parsec squared) to the power n. Observations of distant galaxies have found this relationship to hold consistently, just with varying values of n and A. Schmidt (1959) originally estimated that, n~1, although subsequent studies have found n = 0.9 - 1.7 (Kennicutt, 1998). Kennicutt (1998) attempted to study the global star formation law by analysing observational data from 61 normal spiral galaxies and 36 infrared-selected starburst galaxies. He found that for the 61 normal spiral galaxies, n = 1.29 ± 0.18 or n = 2.47 ± 0.39 depending on the fit used, whilst for the 36 starburst galaxies, n = 1.28 ± 0.08 or n = 1.40 ± 0.13. When the two samples are combined a global star formation law is found with, n = 1.4 ± 0.15 and A = (2.5 ± 0.7) x $10^{-4}$ $M_\odot$ $kpc^{-2}$ $yr^{-1}$. This is commonly known as the Schmidt-Kennicutt law.

The Schmidt-Kennicutt law is an appropriate choice for modelling the SFR because star formation can only occur if there is enough gas, therefore modelling the SFR as a function of gas mass makes physical sense. A disadvantage of using the Schmidt-Kennicutt law is that values of n and A vary considerably in the literature meaning the star formation law is poorly constrained (Kennicutt, 1998). Additionally, Arifyanto et al. (2005) evaluated the Milky Way's star formation history to determine where new stars formed and when over the last 5 billion years (Gyr) of the Galaxy's history. They accomplished this using a sample of M dwarf stars from the solar neighbourhood which they determined to be representative of the entire solar cycle. They demonstrated that a Schmidt-Kennicutt law with, n~1.45, accurately represents the star formation and gas surface density estimated over the last 5 Gyr, agreeing with the results of Kennicutt (1998). Considering these studies it appears as though an exponent of around n = 1.4 is the most accurate, therefore in our study we have utilised a Schmidt-Kennicutt law with n = 1.4 and A = 0.25 $M_\odot$ $pc^{-2}$ $Gyr^{-1}$ (not 2.5 x $10^{-4}$ as this is measured in $M_\odot$ $kpc^{-2}$ $yr^{-1}$ and we want units of $M_\odot$ $pc^{-2}$ $Gyr^{-1}$) to predict the star formation rate from the mass of gas:

$$\psi = aA \left(\frac{M_{gas}}{a}\right)^n \qquad (2)$$

The Schmidt-Kennicutt law relates the SFR surface densities to the surface densities of gas so in order to calculate the SFR, ψ from the gas mass we must divide each quantity by the area of the annulus it represents:

$$a = \pi \left(\left(r + \frac{\Delta r}{2}\right)^2 - \left(r - \frac{\Delta r}{2}\right)^2\right) \qquad (*)$$

Here Δr is the range of each radial bin.

In our GCEM we calculated the SFR at each timestep from the mass of gas at the previous timestep for each radial bin using our Schmidt-Kennicutt law function, ψ. We could then calculate the change in the mass of stars, $\Delta M_s$, which is added to the mass of stars and subtracted from the mass of gas at each timestep and for each radial bin:

$$SFR(\tilde{\imath} + 1, j) = \psi\big(M_{gas}(\tilde{\imath}, j), r(\tilde{\imath}), \Delta r\big)$$

$$\Delta M_s = \big(t(\tilde{\imath} + 1) - t(\tilde{\imath})\big) SFR(\tilde{\imath} + 1, j)$$

$$M_{gas}(i + 1, j) = M_{gas}(\tilde{\imath}, j) - \Delta M_s$$

$$M_{star}(i+1, j) = M_{star}(i, j) + \Delta M_s$$

Here *i* and *j* represent the time and radial bins respectively and all masses are calculated in solar masses, $M_\odot$.

### § 2.3 Initial Mass Function

The initial mass function (IMF) is a function which describes the distribution of stellar masses in a population of stars. We have defined the range of stellar masses to be between 0.08 and 100 $M_\odot$ as used in Kroupa (2001) and for consistency with Gowanlock et al. (2011), although most IMFs now cover stellar masses between 0.1 and 120 $M_\odot$ (Pagel, 2009). The IMF we have chosen to use is that first proposed by Kroupa (1993). This is consistent with other literature (Lineweaver et al. 2004, Gowanlock et al. 2011), although Gowanlock et al. (2011) also uses the Salpeter IMF (Salpeter et al. 1955) in order to test the effects of the two different IMFs on their results. The Salpeter IMF is a simple power law function ($\alpha = 1.35$ and $\xi_n = 0.03$ in (3)), whilst the Kroupa IMF is a two part power law function:

$$\xi(M) = \xi_n \left(\frac{m}{M_\odot}\right)^{-\alpha} \quad (3)$$

$\alpha = 0.3$, $\quad 0.08 \leq m/M_\odot \leq 0.50$, $\quad\quad\quad \alpha = 1.3$, $\quad 0.50 \leq m/M_\odot$

$\xi_n$ is calculated for each alpha by normalising the integrals of the IMF to equal 1 over the entire mass range and using the fact that the IMF is continuous at the boundaries to solve the set of simultaneous equations:

$$\int_{0.08}^{100} \xi(M)\, dm = \int_{0.5}^{100} \xi_0 \left(\frac{m}{M_\odot}\right)^{-1.3} dm + \int_{0.08}^{0.5} \xi_1 \left(\frac{m}{M_\odot}\right)^{-0.3} dm = 1 \quad (A)$$

$$\xi_0 (0.5)^{-1.3} = \xi_1 (0.5)^{-0.3}$$

$$\Rightarrow 2\xi_0 = \xi_1 \quad (B)$$

In our model stellar masses, m are represented in solar mass units, $M_\odot$ and so $m/M_\odot$ can simply be written as M. Equation (A) can be simplified:

$$\Rightarrow -\frac{\xi_0}{0.3}[M^{-0.3}]_{0.5}^{100} + \frac{\xi_1}{0.7}[M^{0.7}]_{0.08}^{0.5} = 1$$

$$\Rightarrow -\frac{\xi_0}{0.3}(100^{-0.3} - 0.5^{-0.3}) + \frac{\xi_1}{0.7}(0.5^{0.7} - 0.08^{0.7}) = 1$$

$$\Rightarrow 3.27\xi_0 + 0.64\xi_1 = 1 \quad (C)$$

By equating equations (B) and (C) we retrieve the two constants:

$\Rightarrow \xi_0 = 0.22 \quad\quad\quad \Rightarrow \xi_1 = 0.44$

It has generally been shown that the Salpeter IMF is a good approximation for masses above 1M$_\odot$ (Larson, 2005) and it has been verified in studies of star formation in spiral galaxies (Kennicutt, 1983). However, Kroupa (2001) and Chabrier (2003) have suggested that at subsolar masses alpha should take a lower value than in the Salpeter IMF thereby producing less low-mass stars and more high-mass stars. Choosing the right IMF is important for studies like this because the number of high-mass stars predicted by a model determines the number of type II supernovas and therefore the fraction of planetary biospheres sterilised by these supernovas. For this reason, we would expect more habitable planets to be predicted by a model using the Salpeter IMF compared with the Kroupa IMF, which is what Gowanlock et al. (2011) found. The flattening of the IMF at subsolar masses is supported by recent observational evidence (Kroupa, 2001, Chabrier, 2003), therefore it is justified to use the Kroupa IMF in our study.

At any point in our simulations where we required only stars of a certain type, the IMF could be integrated to tell us what fraction of a population of stars are in a specific mass range. For example Sun-like (G-type) stars cover the mass range, 0.8 M$_\odot$ ≤ M ≤ 1.2 M$_\odot$ (Strobel, 2010), therefore the fraction of Sun-like stars for an average stellar population is given by:

$$\int_{0.8}^{1.2} \xi(M)\, dm = \int_{0.8}^{1.2} \xi_0 \left(\frac{m}{M_\odot}\right)^{-\alpha} dm = 0.22 \int_{0.8}^{1.2} M^{-1.3}\, dm \approx 0.09$$

### § 2.4 Main Sequence Lifetime

During its life a star fuses elements in its core to make heavier elements. First hydrogen is burned to make helium and then helium is burned to make lithium and so on, creating heavier elements with larger atomic mass each time. When a star is burning hydrogen to form helium it is said to be on its main sequence (MS) (Cosmos, n.d.), the length of time a star spends on its MS is known as its main sequence lifetime (MSL). A star typically spends around 90% of its life on its MS (Cosmos, n.d.) because the Universe and all of the stars and gas within it are predominately composed of hydrogen. A stars MSL is given by the following equation (Cosmos, n.d., Sparke and Gallagher, 2007):

$$\tau_{MS} = \tau_{MS,\odot} \left(\frac{m}{M_\odot}\right)^{-2.5} = 10\, M^{-2.5} \qquad (4)$$

Here $\tau_{MS,\odot}$ = the Sun's MSL ~ 10 Gyr.

As it can be seen from the above equation, a stars MSL increases as stellar mass decreases. The largest stars (O stars) typically have a MSL of only a few million years whilst the smallest stars (M dwarfs) usually have MSLs of over 100 billion years, much longer than the age of the universe (~13.7 Gyr (Cosmos, n.d.)).

In our simulations we assume all stars are on their MS. We do not account for the time between a star coming of its MS and exploding as supernovae as this time is negligibly small in comparison to the lifetime of the star. Additionally, we do not remove G-type stars from our model once they have ended their MS because G stars are very long-lived stars, therefore we would expect the majority of G stars which have been born since the formation of the Galaxy to still be on their MS today.

**§ 2.5 Supernovae**

Supernovae (SNe) are the explosive deaths of stars and are responsible for seeding our Galaxy with all the metals heavier than iron (Sparke and Gallagher III, 2007). The timescales for these explosions varies depending on the type of supernova. For example, Type Ia supernovae (SNIa) occur when a white dwarf in a binary system collapses under its own gravity. If matter is added to a white dwarf by a binary companion, taking its mass above the Chandrasekhar limit of 1.4 $M_\odot$, it can no longer support its own weight and implodes. This heats the interior, triggering nuclear burning, which blows the star apart (Nave, n.d., Sparke and Gallagher III, 2007). Nothing is left; all the iron, nickel, and elements of a similar atomic mass are released back to the interstellar gas. Many of the stars that explode as SNIa do so only at ages of a billion years or more (Sparke and Gallagher III, 2007). Alternatively, stars more massive than 8 $M_\odot$ (Kennicutt, 1984) end their lives by exploding as Type II supernovae (SNII) and release mainly lighter elements with fewer than about 30 neutrons and protons, such as oxygen, silicon and magnesium, back into the interstellar gas. Most of the heavier nuclei such as iron are swallowed up into the remnant neutron star or black hole. These massive stars go through their lives within 100 million years (Myr) (Sparke and Gallagher III, 2007).

In our study we modelled both SNIa and SNII in order to estimate the fraction of stars irradiated by a SNe flux during their lifetime. Gehrels et al. (2003) investigated the effects of SNII on planetary atmospheres and found that the depletion of ozone caused by a nearby SNe flux could cause an extinction event for any land based animal life existing around those planets if the SNe occurred sufficiently close enough. We assume that all stars with mass, $M > 8 M_\odot$ become SNII at the end of their MS and any white dwarf candidate star (0.08 $M_\odot \leq M \leq 8 M_\odot$) has a 1% chance of becoming a SNIa at the end of their MS. This is using the results of Pritchet et al. (2008) who estimated that ~1% of white dwarfs become SNIa after their MS has ended, independent of mass.

In our simulation we created two functions to work out the rate of SNe at each timestep, t. To do this we needed to see the rate at which SNe progenitor stars were being born at time, $t-\tau_{MS}$ (where $\tau_{MS}$ is the MSL of the SNe progenitor star), this was achieved by interpolating the SFR at time, $t-\tau_{MS}$. In order to retrieve the rate of SNe we then integrated this multiplied by the IMF over the range of SNe progenitor stars and in the case of SNIa multiplied this value by 0.01:

$$\int_{8}^{100} \frac{\xi(M)}{M} SFR(t - \tau_{MS}(M))\, dM \qquad \text{(SNII)}$$

$$0.01 \int_{0.08}^{8} \frac{\xi(M)}{M} SFR(t - \tau_{MS}(M))\, dM \qquad \text{(SNIa)}$$

Here the IMF is divided by mass, M so that it represents the number of stars as opposed to the mass of stars in the required mass range. We computed the integrals using the trapezium method because originally we planned to incorporate the infall rate of extragalactic gas into our model, which would have made our SFR function much more chaotic and therefore integrating using the trapezium method would have been the most accurate option.

In our main program we then multiplied the SNe rates by t(i+1)-t(i) in order to retrieve the number of SNII and SNIa at each timestep and for each radial bin and saved these values to two matrices SNII(i,j) and SNIa(i,j) respectively. In order to estimate how many potentially habitable star systems (PHSS) have been sterilised by nearby SNe we calculated the unsterilized area of each annulus for stars born at each timestep. We did this by iterating over all SNe which took place more than 2.25 Gyr after the stars had formed and applied this equation at each timestep and for each radial bin:

$$survarea(y,j) = survarea(y,j) - \frac{Sx * survarea(y,j)}{totarea(j)}$$

Here totarea(j) is the area of each annulus calculated using equation (*) and initially survarea(y,j) is set to equal totarea(j) for each timestep and for each radial bin. $Sx = S1 = 64\pi$ pc$^{-2}$ and $Sx = S2 = 361\pi$ pc$^{-2}$ are the predicted sterilisation areas of SNII and SNIa respectively. S1 is calculated using the results of Gehrels et al. (2003) who found that an average SNII would cause an extinction event on planets located ≤ 8 pc away. Whilst S2 was calculated using figure 3 from Gowanlock et al. (2011) who estimated the average sterilisation distance for SNIa to be ~ 19 pc. For stars which were born early in the Galaxy's history, we assume that they will still be around in all future timesteps, therefore the unsterilized area must be reduced by SNe which occur in all future timesteps, represented by y. Finally we have used periodic boundary conditions which is a reasonable assumption as S1 and S2 are much less than the area of each annulus.

Gowanlock et al. (2011) states that the Earth's ozone layer was only sufficiently developed ~2.3 billion years ago, therefore if a SNe flux had irradiated the Earth before this time it would unlikely have had much of an effect on our own evolution. The time at which SNe sterilisations occur in a stars life is therefore very important to consider in studies of this nature which is why we have only reduced the unsterilised area for SNe which occurred in timesteps more than 2.25 Gyr after the stars formed. This means that SNe in our model do not have a sterilising effect on stars born in the same timestep, however, they do for stars born 2.25 Gyr or more before that timestep.

### § 2.6 Metallicity

During its life, the Galaxy turns gas into stars. Each star burns hydrogen and helium to form heavier elements, which are returned to the interstellar gas at the end of its life (Sparke and Gallagher III, 2007). This process of fusing lighter nuclei to produce heavier nuclei, is known as nucleosynthesis and without it life on Earth could not exist (Christian, 2012). In Astronomy any element heavier than Hydrogen and Helium is known as a "metal". The proportion of these metals that an astronomical object contains is known as its "metallicity". An objects metallicity is usually expressed as the ratio of the mass of iron to the mass of hydrogen compared with the Sun and can be calculated with the following formula:

$$[Fe/H] = log_{10}\left(\frac{M_{Fe}}{M_H}\right)_{star} - log_{10}\left(\frac{M_{Fe}}{M_H}\right)_{sun} \qquad (5)$$

Here $M_{Fe}$ and $M_H$ are the mass of iron and mass of hydrogen per unit of volume respectively (Sanders, 2012).

In our simulation we utilise data from Raiteri et al. (1996) and Iwamoto et al. (1999) which define estimates for the total mass ejected, Mej, the mass of iron ejected, MejZ, and the mass of oxygen ejected, MejO, by both SNIa and SNII, in order to estimate the increase in metallicity with time as metals are returned to the interstellar gas by SNe explosions. We do not account for other processes which enrich the gas with metals such as the stripping of surface gas from asymptotic giant branch stars via stellar superwinds or the return of metals to the gas from normal nucleosynthesis in stars which do not explode as SNe (Sparke and Gallagher, 2007). This is a reasonable estimation as these processes contribute significantly less material to the gas than SNe explosions do.

$$Mej_{SNII} = \left(t(i+1) - t(i)\right) \int_{8}^{100} \frac{\xi(M)}{M} SFR(t - \tau_{MS}(M)) \, x \, 0.7682 M^{1.056} \, dM$$

$$MejZ_{SNII} = \left(t(i+1) - t(i)\right) \int_{8}^{100} \frac{\xi(M)}{M} SFR(t - \tau_{MS}(M)) \, x \, (2.802 \, x \, 10^{-4}) M^{1.864} \, dM$$

$$MejO_{SNII} = \left(t(i+1) - t(i)\right) \int_{8}^{100} \frac{\xi(M)}{M} SFR(t - \tau_{MS}(M)) \, x \, (4.586 \, x \, 10^{-4}) M^{2.721} \, dM$$

$$Mej_{SNIa} = 0.01 * \left(t(i+1) - t(i)\right) \int_{0.08}^{8} \frac{\xi(M)}{M} SFR(t - \tau_{MS}(M)) \, x \, 1.38 \, dM$$

$$MejZ_{SNIa} = 0.01 * \left(t(i+1) - t(i)\right) \int_{0.08}^{8} \frac{\xi(M)}{M} SFR(t - \tau_{MS}(M)) \, x \, 0.626 \, dM$$

$$MejO_{SNIa} = 0.01 * \left(t(i+1) - t(i)\right) \int_{0.08}^{8} \frac{\xi(M)}{M} SFR(t - \tau_{MS}(M)) \, x \, 0.143 \, dM$$

These SNe yields were calculated using our two SNe functions. The mass ejected by both types of SNe is added to the mass of gas and subtracted from the mass of stars at each timestep and for each radial bin. Whilst the mass of iron and oxygen ejected by both types of SNe are used to calculate the fraction of gas that is composed of iron and oxygen at each timestep and for each radial bin which are stored in the matrices, Z and O respectively:

$$M_{gas}(i+1,j) = M_{gas}(i+1,j) + Mej_{SNII} + Mej_{SNIa}$$

$$M_{star}(i+1,j) = M_{star}(i+1,j) - Mej_{SNII} - Mej_{SNIa}$$

$$Z(i+1,j) = Z(i,j) \frac{M_{gas}(i,j)}{M_{gas}(i+1,j)} + \frac{MejZ_{SNII} + MejZ_{SNIa}}{M_{gas}(i+1,j)}$$

$$O(i+1,j) = O(i,j) \frac{M_{gas}(i,j)}{M_{gas}(i+1,j)} + \frac{MejO_{SNII} + MejO_{SNIa}}{M_{gas}(i+1,j)}$$

Equation (5) can then be used to work out the Galaxy's metallicity profiles in terms of iron and oxygen using the matrices, Z and O:

$$FeH(i+1,j) = \log_{10}(Z(i+1,j)) - \log_{10}\left(\frac{1.17 \, x \, 10^{-3}}{0.706}\right)$$

$$OxH(i+1,j) = \log_{10}(O(i+1,j)) - \log_{10}\left(\frac{9.59 \, x \, 10^{-3}}{0.706}\right)$$

Here the Sun's iron, oxygen and hydrogen abundances are taken as being 1.17 x 10$^{-3}$ M$_\odot$, 9.59 x 10$^{-3}$ M$_\odot$ and 0.706 M$_\odot$ respectively as estimated by Anders and Grevesse (1989).

## § 2.7 Calculating the Galactic Habitable Zone

Once all of these properties of the Galaxy had been incorporated into our GCEM we could work out our GHZ as being the region containing all the PHSS born since the formation of the Galaxy. We define a PHSS as one that has a high enough metallicity for Earth-sized planet formation to take place ([Fe/H] > -1 using the results of Johnson and Hui (2012) who estimated that above this critical metallicity Earth-sized planets can form) and has remained free from SNe sterilisations long enough for land-based animal (complex) life to evolve (~3.8 Gyr here on Earth (Gowanlock et al. 2011)).

The number of G stars formed at each timestep and for each radial bin was calculated using the SFR and by integrating the IMF between the mass range of G-type stars:

$$G(i+1, j) = \left(t(\tilde{i}+1) - t(\tilde{i})\right) * SFR(i+1, j) \int_{0.8}^{1.2} \frac{\xi(M)}{M} dM$$

The number of G stars formed which survived SNe sterilisations, Gm could then be calculated using:

$$Gs(i+1, j) = G(i+1, j) * \frac{survarea(i+1, j)}{totarea(j)}$$

By removing any stars whose metallicity is less than 1 we retrieved the number of stars which have survived SNe sterilisations and which have a high enough metallicity for Earth-sized planet formation, Gm. Finally we removed any stars which had formed in the last 3.8 Gyr as these stars may not have had long enough for complex life to evolve on planets surrounding them yet. Consequently, we had estimated the number of PHSS formed at each timestep and for each radial bin, Ga, which make up our GHZ.

## § 2.8 The Age Distribution of Potential Intelligent Life in our Galaxy

Finally we worked out the age distribution of the PHSS in our model. The age distribution of intelligent life could then be inferred from this using the age of our Sun and the estimated timescales for our own evolution as a proxy. The age distribution of the PHSS was calculated using a weighted mean:

$$agedis(j) = \frac{\sum(Ga(:,j) * t')}{\sum Ga(:,j)}$$

$$agedis = 14 - agedis$$

Here Ga(:,j) is an array which represents the number of PHSS formed at each timestep and j represents each radial bin. The second line converts from average formation time to average age of the stars. By comparing the average ages of PHSS in our Galaxy to the age of the Solar System (~4.55 Gyr) we are able to see how old potential intelligent civilisations elsewhere in the Galaxy may be in comparison to us (assuming that the evolution of intelligent life on other planets occurs under the same timescales as it has done here).

## § 2.9 Model adjustment and implementation

When running our simulations we found that the SNIa rate was approximately an order of magnitude higher than the SNII rate, when in fact observational data has shown that the SNIa rate is about an order of magnitude lower than the SNII rate (Pagel, 2009). To correct for this we integrated the IMF by number (equation (1) divided by mass, M (Kroupa, 2001)) over the range of SNII (8 $M_\odot$ ≤ M ≤ 100 $M_\odot$) multiplied by 0.1 and equated this to the IMF by number over the range of white dwarf candidate stars (0.08 $M_\odot$ ≤ M ≤ 8 $M_\odot$) multiplied by β. Where β is the fraction of white dwarfs which become SNIa, which we calculated to be, β≈6.3x10$^{-4}$. When we applied this new fraction β in our model we still had more SNIa than SNII. The reason for this was unknown and was probably due to an error in our code. However, because of time constraints we were unable to correct this, therefore we removed the SNIa function described in section 2.5 and instead defined the number of SNIa as being equal to the number of SNII multiplied by 0.1 at each timestep and for each radial bin. This may be inaccurate because although this is the observed ratio between SNIa and SNII at the present day it may not have been in the past. Additionally, we are not accounting for the fact that SNIa explode on much longer timescales than SNII, seeding the Galaxy with metals at different times in the Galaxy's history. Therefore, we may have overestimated the number of SNIa and consequently the metallicity during the first few billion years of the Galaxy's history.

Now that we had set up our GCEM we could test our model for different values of $r_d$ and $\Sigma_0$ in equation (1). As previously stated the scale length, $r_d$, is estimated to be between 2000-4500 pc at the present day (Naab and Ostriker, 2006, Sparke and Gallagher, 2007 and Binney and Tremaine, 2008). When the Galaxy formed we would expect the scale length to have been smaller, this is because the Galaxy is predicted to have undergone "inside-out" formation (Prantzos, 2008, Gowanlock et al. 2011), where initially star formation occurs preferentially in the inner Galaxy but decreases with time. For this reason we chose to test our model with values of $r_d$ between 1000-3000 pc. Our GCEM is what is known as a "closed box" model, where all the stars and gas in the model remain in the simulated Galaxy nothing enters or leaves it. This is convenient as it means we can equate estimates for the total mass (gas + stars) of the Galaxy at the present day, $M_{tot}$ to the total gas mass of the Galaxy at its formation (as we assume there are initially no stars just gas), which is defined by equation (1). By integrating (1) over the entire Galaxy we found that $M_{tot}$ = 2π*$r_d$*$\Sigma_0$. Pagel (2009) states that $M_{tot}$ = 7.7x10$^{10}$ $M_\odot$ so $r_d$*$\Sigma_0$ = 7.7x10$^{10}$/2π ≈ 1.225x10$^{10}$ $M_\odot$. Therefore, using this relation we could work out the corresponding $\Sigma_0$ values for each of the $r_d$ values we tested.

When testing our program for different values of $r_d$ and $\Sigma_0$ it soon became apparent that for certain initial conditions more gas was being turned into stars than existed in the first place thereby making the gas mass in the centre of the Galaxy negative after the very first timestep. The reason for this can be seen by looking at equation (2), the exponent of the gas mass surface density is 1.4 therefore we would expect a sufficiently large value for the initial gas mass to result in an even larger SFR causing a negative gas mass in the next timestep. The exponent, n=1.4 is fairly well defined by the Schmidt-Kennicutt relation because this value represents the gradient of the gas-SFR surface density plot which is estimated from observational data, it is therefore unjustified to alter it significantly. The other constant A on the other hand is less well constrained, dictating the intercept of the gas-SFR surface densities plot which could in fact be higher or lower than Kennicutt (1998) originally calculated due to systematic observational errors. Additionally, in our model we do not account for the expulsion of gas caused by SNe which would have the effect of spreading out the gas and therefore reducing the SFR. For these reasons it is justified to use a smaller value of A in our model, therefore we also varied A between 0.01 - 0.2 $M_\odot$ pc$^{-2}$ Gyr$^{-1}$ when testing our model.

After testing our program for different values of $r_d$, $\Sigma_0$ and A we found that the combination of these values which produced present day gas and star surface densities in the Solar neighbourhood in the ratio closest to what has been predicted by other studies were, $r_d$=2000 pc, $\Sigma_0$=6.125x10$^6$ M$_\odot$ pc$^{-2}$ and A=0.05 M$_\odot$ pc$^{-2}$ Gyr$^{-1}$. The surface densities of stars and gas varies considerably between authors. The surface density of gas near the Sun is estimated to be 10 M$_\odot$ pc$^{-2}$ (Sparke and Gallagher, 2007), 7-14 M$_\odot$ pc$^{-2}$ (Pagel, 2009) and 13 M$_\odot$ pc$^{-2}$ (Binney and Tremaine, 2008). Whilst the surface density of stars is estimated to be 25-40 M$_\odot$ pc$^{-2}$ (Sparke and Gallagher, 2007), 45 M$_\odot$ pc$^{-2}$ (Pagel, 2009) and 36 M$_\odot$ pc$^{-2}$ (Binney and Tremaine, 2008). The ratio of the surface densities of stars to gas is therefore ~3:1. Our chosen model predicts the surface density of stars to be ~10 M$_\odot$ pc$^{-2}$ and the gas surface density to be ~4 M$_\odot$ pc$^{-2}$ near the Sun (taking the Sun's position to be 8 kpc (Binney and Tremaine, 2008)). Our surface densities are approximately a factor of 3 lower than estimated previously suggesting that our model may not represent the Milky Way as well as these previous studies have. However, this is unsurprising because of the various galactic properties that we could not model because of time constraints and limited computing power (the vertical component of the disk, infall rate, radial gas inflows, orbits of stars etc.).

# § 3 Results

## § 3.1 Supernovae

As discussed in section 2.5 we modelled SNIa and SNII in order to estimate how many G stars have been sterilised by nearby SNe in the Galaxy's lifetime. Previous research has estimated that the fraction of planets sterilised by nearby SNe is much higher towards the centre of the Galaxy and within and close to the midplane (Gowanlock et al. 2011). The probability of a star born at different times since the Galaxy's formation and at different radial distances from the galactic centre remaining unsterilised by nearby SNe over the entire lifetime of the Galaxy is presented in figure 1. We found that stars formed farther from the galactic centre and at later times in the Galaxy's history are more likely to remain unsterilised. This is because the SFR decreases with time and distance from the galactic centre meaning less SNe are produced. Additionally, the density of stars decreases exponentially as you move from the galactic centre towards the outskirts of the Galaxy which means that SNe result in fewer sterilisations towards the outskirts of the Galaxy as stars are located farther apart. According to our model the Sun was born in a region where the probability of it remaining unsterilised was very low (~4%), supporting the possibility that a nearby SNe may have caused one or more of the five mass extinctions the Earth has experienced during its history, which has been discussed previously (Ellis and Schramm, 1995). All stars born in the last 2.25 Gyr are assumed to have remained unsterilised because of the condition that an ozone layer has to be present for an extinction event to take place (discussed in section 2.5).

Overall we found that ~96% of G-type stars were sterilised by a nearby SNe at some point in the Galaxy's history, with our model estimating that ~90% of G-type stars at the Sun's radial position have been sterilised. We disagree with the conclusion of Gowanlock et al. (2011) that a major increase in SNe sterilisation distances is unlikely to significantly reduce the habitability of the Galaxy. This is because although when the sterilisation area of each SNe is doubled, we found that ~97% of stars are sterilised and ~91% of stars in the solar neighbourhood are sterilised - only a 1% increase on previous figures, this resulted in the total number of PHSS that our model predicted decreasing by over 50%, from ~6.2 million to ~2.9 million. This suggests a major reduction in the habitability of the Galaxy.

The rate of SNe explosions in our Galaxy is very uncertain and values vary considerably in the literature. The fact that they only occur every 10-100 years in our Galaxy (Ellis and Schramm, 1995) coupled with the fact that many go undetected due to absorption by dust in the disk (Gowanlock et al. 2011), makes them very difficult to study. Additionally, our insight from observing them in other galaxies is limited. This is because the large distances involved means any measurements obtained will have a low signal to noise ratio and consequently large uncertainties. Our model estimated the SNIa and SNII rate over the past Gyr to be ~$3.62 \times 10^{-3}$ and ~$3.62 \times 10^{-2}$ respectively.

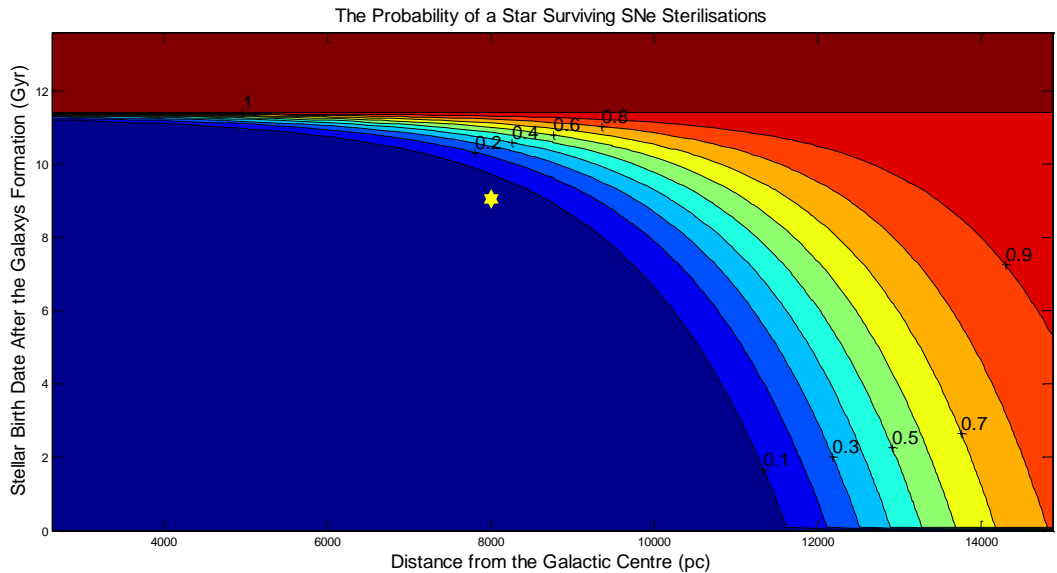

Figure 1: The probability of a star remaining unsterilised until the present day as a function of radial distance from the galactic centre and birth date since the formation of the Galaxy. The yellow star represents the Sun's location (8 kpc) and birth date (9.05 Gyr after the Galaxy's formation). Stars born at larger radial distances and at later times in the Galaxy's history were the most likely to survive SNe sterilisations. This is because the SNe rate decreases with time and the stellar density decreases with radial distance from the galactic centre.

### § 3.2 Metallicity

We investigated how the Metallicity of the interstellar gas in terms of the iron and oxygen abundances changed with time and radial position from the galactic centre. Our results are displayed in figures 2 and 3. We found that the metallicity of the Galaxy in terms of both iron and oxygen increased with time and proximity to the galactic centre. Soon after the Galaxy's formation the overall metallicity in terms of both iron and oxygen increased rapidly, with the increase slowing with time, this can be seen in figure 3. We estimate that the metallicity of the gas at the Sun's radial position when it was born ~4.55 billion years ago was ~0.28 and ~0.79 in terms of iron and oxygen respectively. Clearly we have overestimated the metallicity (especially for oxygen), as these should be 0 by definition (see equation (5)). However, this is probably because we assumed that the SNIa rate was 10 % of the SNII rate at all times in the Galaxy's history, meaning we may have overestimated the number of SNIa early in the Galaxy's history, resulting in the metallicity being overestimated as discussed previously.  Overall the metallicity abundances in terms of oxygen are higher than iron by ~0.5 at all times and radial positions. This is to be expected as SNII release a lot more oxygen than iron and are 10 times more frequent than SNIa in our model which release more iron than oxygen.

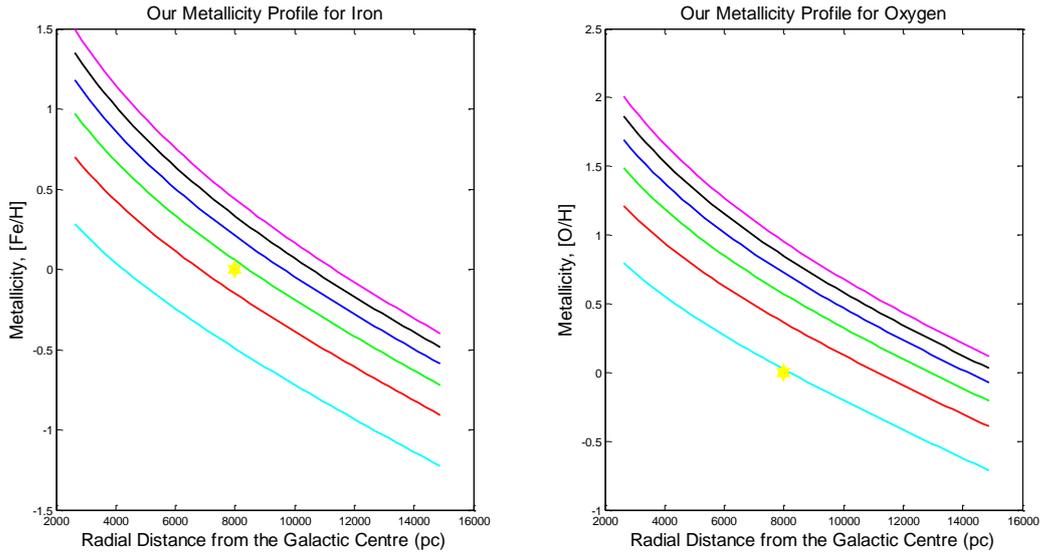

Figure 2: The metallicity profiles for iron and oxygen as predicted by our model, shown as a function of radial distance from the galactic centre. The metallicity is calculated as the logarithm of the ratio of the amount of iron (or oxygen) to hydrogen in the interstellar gas relative to the Sun. The coloured lines represent the metallicity at different times after the Galaxy's formation from 2 Gyr (bottom line) to 12 Gyr (top line) in 2 Gyr increments. The yellow star represents the Sun's radial position (8 kpc) and metallicity, which is 0 by definition. The metallicity is always highest near the galactic centre because of the higher SFR there and the metallicity in terms of oxygen is always ~0.5 times higher than the metallicity in terms of iron. Our metallicity profile for iron can be directly compared with figure 11 in Naab and Ostriker (2006) (see appendix B).

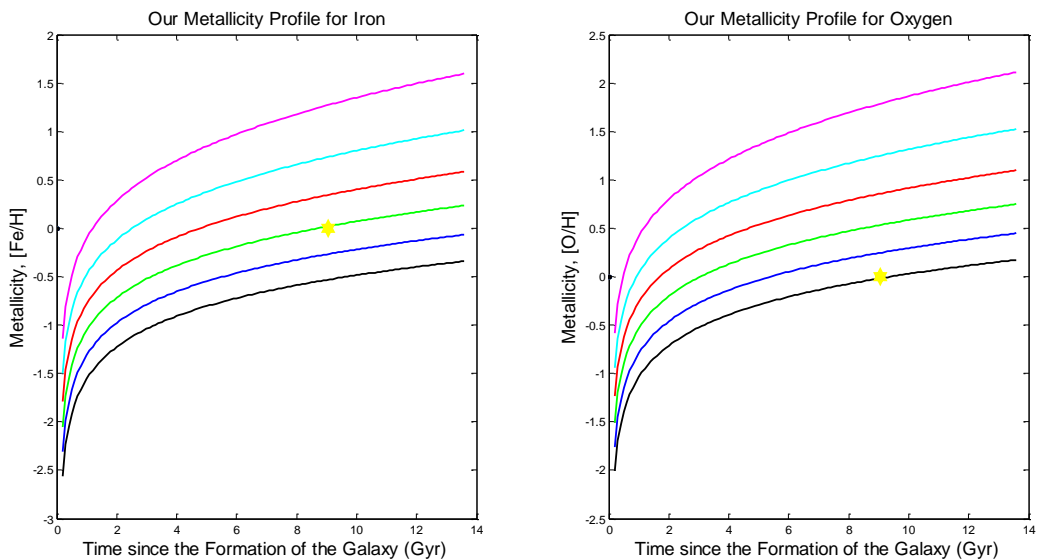

Figure 3: the metallicity profiles for iron and oxygen as predicted by our model, shown as a function of time since the formation of the Galaxy. The coloured lines represent the metallicity at different radial positions from the centre of the Galaxy from ~2.5 kpc (top line) to ~15 kpc (bottom line) in ~2.5 kpc increments. The yellow star represents the Sun's formation time (~9.05 Gyr) and metallicity, which is 0 by definition. The metallicity increases rapidly soon after the Galaxy's formation with the increase slowing with time. This is because initially there is a very high SFR which declines with time. Our metallicity profile for iron can be directly compared with figure 1 in Lineweaver et al. (2004) (see appendix C).

**§ 3.3 The Galactic Habitable Zone**

A GHZ was calculated using the criteria described in section 2.7. Overall our model estimated there to be ~$2.3 \times 10^9$ G-type stars in our Galaxy. We estimated that ~10% of these stars did not have a high enough metallicity for Earth-sized planet formation to take place ([Fe/H] > -1 (Johnson and Hui, 2012)), ~6% of these stars are not old enough for complex life to have evolved on potentially habitable planets surrounding them yet (< 3.8 Gyr old) and ~96% of these star systems have been sterilised by nearby SNe, as stated previously. The number of G stars formed per Gyr, the number of unsterilised G stars formed per Gyr, the number of unsterilised G stars with a high enough metallicity formed per Gyr and finally the number of unsterilised G stars with a high enough metallicity and which are old enough for complex life to have evolved formed per Gyr are presented in figure 4, as a function of formation time summed over the entire Galaxy.

After removing all the G-type stars which did not meet the criteria stated above we were left with ~6.2 million PHSS in which complex life and consequently intelligent life could have evolved. These PHSS make up our GHZ shown in figure 5 as a function of radial position and time and normalised by dividing by the total number of PHSS so that the contour lines illustrate the percentage of PHSS born in each region. The Sun is much younger and located much closer to the galactic centre than the majority of PHSS. The flat line at the top of the GHZ reflects the condition that a star has to be over 3.8 Gyr old in order to be considered a PHSS. The curved boundary on the left of the GHZ is caused by too many SNe sterilisations. This is because at earlier times in the Galaxy's history more stars are sterilised due to the higher SNe rate and closer to the galactic centre there are more sterilisations due to the higher SNe rate and stellar density. The final boundary at the bottom right of the GHZ is caused by stars having a metallicity too low for Earth-sized planet formation. This is because stars born farther from the galactic centre and at earlier times in the Galaxy's history have a lower metallicity and therefore do not form Earth-sized planets under our model's assumptions. Overall our GHZ encompasses the region, 7-14 pc from the galactic centre and contains stars formed between 11 and 3.8 billion years ago. The highest concentration of PHSS are located 9-11 kpc from the galactic centre and were formed between 3.8-6 billion years ago.

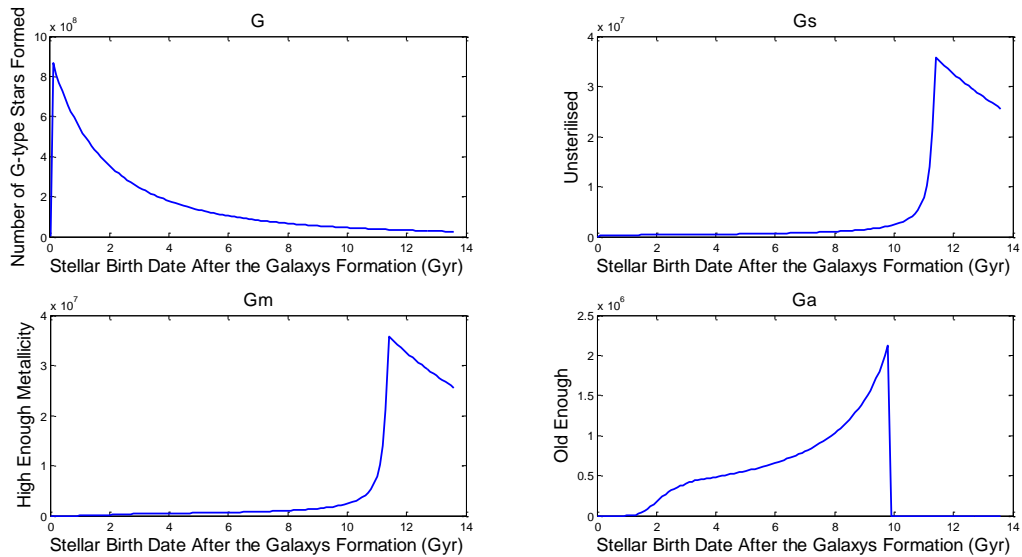

Figure 4: The number of G stars formed per billion years (top left), which remained unsterilised by nearby SNe (top right), had a high enough metallicity for Earth-sized planet formation (bottom left) and are old enough for complex life to have evolved on potentially habitable planets surrounding them (bottom right). The majority of stars have been sterilised by a nearby SNe at some point during their lifetime. The number of PHSS formed per billion years (bottom right) has been slowly increasing over the last 10 Gyr. Therefore, we would expect many more PHSS to be born over the next few billion years.

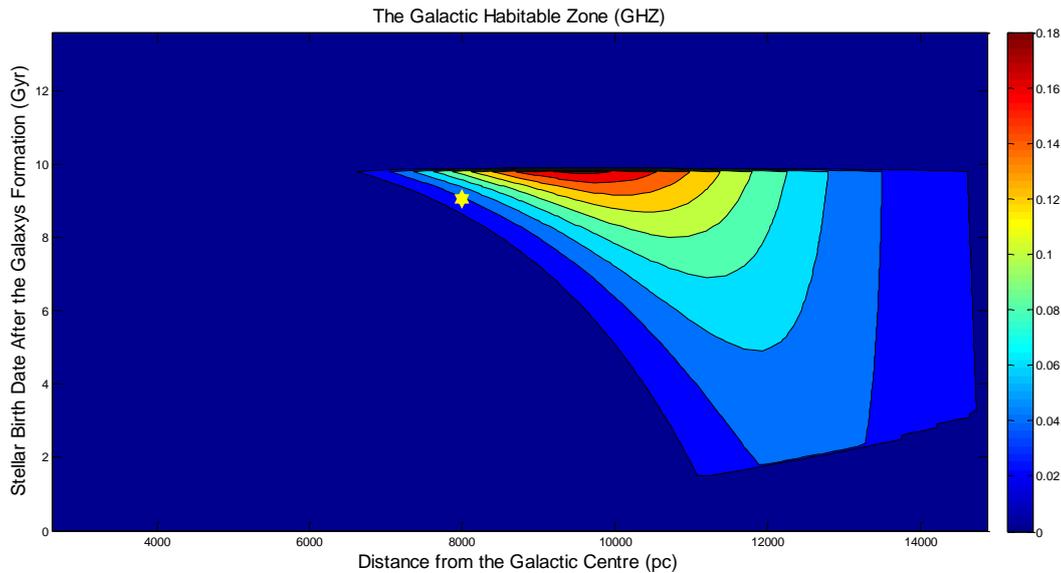

Figure 5: The Galactic Habitable Zone (GHZ) as a function of radial distance from the galactic centre and formation time. The GHZ contains all PHSS formed over the lifetime of the Galaxy. The yellow star represents the Sun's formation time (~9.05 Gyr) and radial position (8 kpc). The number of PHSS born in each region has been divided by the total number of PHSS so that the contour lines represent the percentage of PHSS being born in each region. The majority of PHSS are much older and located much farther from the galactic centre than the Sun.

## § 3.4 The Age Distribution of Potential Intelligent Life in the Milky Way

Finally we were able to estimate the age distribution of PHSS in the Milky Way. Our results are displayed in figure 6 which shows the average ages of PHSS at different radial distances from the galactic centre. It is clear that the majority of PHSS are a lot older than our Solar System. In fact we predict that ~77% of PHSS are older than our Sun with their average age being ~7.68 Gyr, ~3.13 Gyr older than our Sun. Overall the average age of a PHSS is ~6.97 Gyr, ~2.42 Gyr older than the Sun. Because the average age of PHSS is billions of years older than our Sun we would expect any intelligent life in the Galaxy to be incredibly more advanced than we are as the majority of other star systems have had a much longer time for intelligent life to develop and evolve. At the Sun's radial position the average age of a PHSS is ~4.79 Gyr, slightly older than the age of our Solar System, ~4.55 Gyr, suggesting that if we discover intelligent life in the solar neighbourhood they may be a lot more evolved than we currently are (~240 Myr older on average). However, this is highly speculative as we are assuming that intelligent life on other planets evolves under the same timescales as it has done here, which is an unlikely but necessary condition.

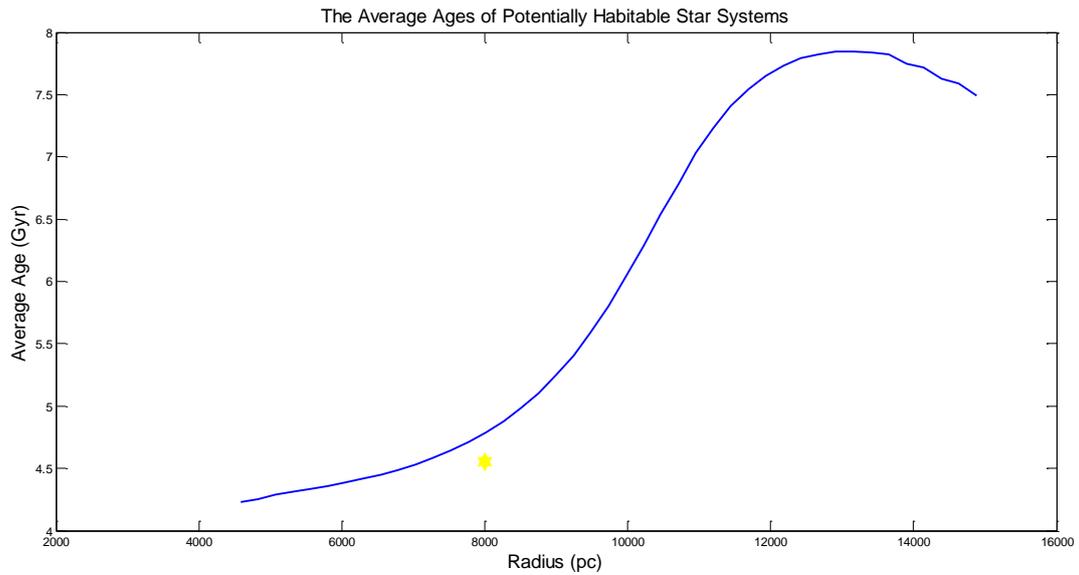

Figure 6: The age distribution of potentially habitable star systems (PHSS) as a function of radial distance from the galactic centre. The yellow star represents the Sun's age (~4.55 Gyr) and radial position (8 kpc). We find that PHSS located closer to the galactic centre than the Sun's radial position are on average younger than the Sun, whilst PHSS located farther from the galactic centre are on average much older than the Sun. The oldest PHSS are located at ~13 kpc where their average age is ~7.8 Gyr, much older than the Sun.

# § 4 Discussion

## § 4.1 Comparison of Results with Previous Studies

### § 4.1.1 Additional Properties of our Simulated Galaxy

We have stated that our simulated Galaxy may not be the best representation of the Milky Way because of its inability to reproduce present day surface densities for the gas and stars at the Sun's radial position close to what other models have predicted, but what about the other characteristics of our Galaxy? Sparke and Gallagher (2007) report that there is ~5-10x$10^9$ M$_\odot$ and 6x$10^{10}$ M$_\odot$ of gas and stars respectively in our Galaxy's disk. Whilst Binney and Tremaine (2008) state that there is total mass of stars and gas of ~4.5x$10^{10}$ M$_\odot$. Our model estimates that there is ~2.35x$10^9$ M$_\odot$ and 1.97x$10^{10}$ M$_\odot$ of gas and stars respectively in our Galaxy's disk. These values are approximately a factor of 3 lower than the values estimated by Sparke and Gallagher (2007) and approximately a factor of 2 lower than the value estimated by Binney and Tremaine (2008) although our gas fraction is consistent with that reported by Pagel (2009), ~0.1. Additionally, we estimate the global time-averaged SFR and local time-averaged SFR surface density to be ~1.78 M$_\odot$yr$^{-1}$ and 0.89 M$_\odot$pc$^{-2}$Gyr$^{-1}$ respectively. Again these estimates are approximately a factor of 3 lower than previous estimates, with Pagel (2009) reporting a global time-averaged SFR of 5-7 M$_\odot$yr$^{-1}$ and a local time-averaged SFR surface density of 3-4.5 M$_\odot$pc$^{-2}$Gyr$^{-1}$. What this suggests is that although our simulated Galaxy may not perfectly resemble the Milky Way it could be considered to represent a scaled down version of the Milky Way approximately one third of the size. This shouldn't dissuade us from making conclusions about the habitability and age distribution of PHSS in the Milky Way because our model predictions are in the same proportions as previous studies, just one third the magnitude.

### § 4.1.2 Supernovae

Gowanlock et al. (2011) estimated that ~64-73% of stars were sterilised by SNe at some point in the Galaxy's history depending on which model they used. Additionally, they estimated that at the Sun's radial position, r=8kpc, ~47-64% of stars were sterilised. These values are significantly lower than our model which predicted that ~96% and ~90% of G stars were sterilised for the entire Galaxy and at the Sun's radial position respectively. A possible reason for this might be because Gowanlock et al. (2011) modelled the vertical component of the Galaxy whereas we did not. This is because a group of stars which appear close to each other (and therefore vulnerable to SNe sterilisations) in terms of radial position in our model, may in fact be far away from one another once their vertical positions are taken into consideration, therefore reducing the fraction of SNe sterilisations. For this reason our study may have significantly overestimated the fraction of SNe sterilisations, suggesting that the Galaxy may in fact be more habitable than we have predicted here. This is especially true towards the centre of the Galaxy where the stellar density and therefore the fraction of SNe sterilisations is greatest.

Gowanlock et al. (2011) estimated the global SNIa and SNII rates over the past Gyr to be ~6.6x$10^{-3}$ yr$^{-1}$ and ~3.7-4.4x$10^{-2}$ yr$^{-1}$ respectively (for their two models also using the Kroupa IMF). Our model estimated the SNIa and SNII rate over the past Gyr to be ~3.62x$10^{-3}$ yr$^{-1}$ and 3.62x$10^{-2}$ yr$^{-1}$ respectively. Our SNII rate is in good agreement with Gowanlock et al. (2011), however, our SNIa rate is about a factor of two lower. This is probably due to the fact that we assumed the SNIa rate was 10% of our SNII rate instead of modelling SNIa independently. Therefore, our model may have slightly underestimated the number of SNIa and subsequently the fraction of sterilisations in the last billion years.

### § 4.1.3 Metallicity

Our metallicity profile for iron in figure 1 can be directly compared with figure 11 of Naab and Ostriker (2006) (appendix B). Our metallicity profile follows the same general trend as in Naab and Ostriker (2006), with metallicity increasing with time and decreasing with radial distance from the galactic centre. However, their plot indicates a steeper metallicity gradient at earlier times in the Galaxy's history whereas our plot has a relatively constant metallicity gradient at all times. Additionally, our metallicities reach higher values especially near the galactic centre where our model predicts metallicities of ~1.5 compared with their model which has a maximum metallicity of ~0.5. Our metallicity profile for iron from figure 3 can be directly compared with figure 1 of Lineweaver et al. (2004) (appendix C). Again, our metallicity profile follows largely the same trend with metallicity increasing with time and decreasing with distance from the galactic centre. Our metallicity profile extends to a value of ~1.5 for the inner Galaxy, whilst Lineweaver et al. (2004) metallicity profile only extends up to ~0.5. Additionally, their metallicities decreased slightly ~10 billion years ago, especially near the galactic centre. However, this is likely due to the fact that they included two overlapping episodes of accretion in their model, something which we did not include in our model.

### § 4.1.4 The Galactic Habitable Zone and the Age Distribution of Complex Life

Our results indicate a different GHZ than previous studies (Lineweaver et al. 2004, Gowanlock et al. 2011). For instance, Lineweaver et al. (2004) found that a habitable zone first emerged ~8 billion years ago (68% contour) which expanded with time as metallicity spread to the outer Galaxy and the SNe rate decreased, but always remained centred at 8 kpc. Whereas Gowanlock et al. (2011) found that the inner Galaxy was the most habitable integrated over all epochs of the Galaxy's history, with the highest number and fraction of habitable planets being located between 2.5-5 kpc and within and surrounding the midplane. Additionally, Gowanlock et al. (2011) predicted that ~0.3% of stars may host a non-tidally-locked HZ planet. Given the result that no planets orbiting stars in the G star mass range are expected to experience tidal locking (Gowanlock et al. 2011) and using the results of Petigura et al. (2013) who estimated from observational data of exoplanets that ~22% of Sun-like stars host a HZ planet, we estimate that ~0.06% $(((6.2 \times 10^{6}) \times 0.22 \times 100) / (2.3 \times 10^{9}))$ of G stars in the Galaxy may host a non-tidally locked HZ planet. This figure is ~5x lower than the value estimated by Gowanlock et al. (2011), however, it is worth noting that their results apply to F, G, K and M stars while ours only apply to G stars.

In general the GHZ generated in the current study is located farther from the galactic centre and first appeared earlier in the Galaxy's history than previous research has predicted. This is probably due to the fact that we didn't model planet formation, assuming that any star with a metallicity above the critical metallicity of [Fe/H] > -1 had the same capability of forming Earth-sized habitable planets. This is contrary to Lineweaver et al. (2004) and Gowanlock et al. (2011) who assumed that the probability of forming habitable planets increased with metallicity (until too high a metallicity meant that a hot Jupiter would form which they assumed would destroy any habitable planets also present). Our method seems justified based on the recent observational analysis of Buchhave et al. (2012) which suggests that there is no special requirement of enhanced metallicity for terrestrial planet formation. If we had assumed that the probability of forming habitable planets increased with metallicity as these two previous GHZ studies (Lineweaver et al. 2004, Gowanlock et al. 2011) did, we would expect our GHZ to be restricted in the outer Galaxy and at earlier times when the metallicity was lower and enhanced in the inner Galaxy and at later times when the metallicity was higher, therefore producing a GHZ more similar to Lineweaver et al. (2004).

Our results are similar to Lineweaver et al. (2004) in that they predicted that ~77% of stars that might harbour complex life in the Galaxy are older than the Sun compared with Lineweaver et al. (2004) who predicted this fraction to be ~75%. However, they estimated the average age of these stars to be ~1 Gyr older than the Sun, whereas our model predicts the average age of these older stars to be ~3.13 Gyr older than the Sun. This inconsistency between the two models is likely caused by the differing ways in which we modelled metallicity, as discussed above. This is because older stars have lower metallicities which would reduce the likelihood of the formation of habitable planets under Lineweaver et al. (2004) assumptions, however, such stars will definitely form habitable planets under our studies assumptions providing that their metallicity is greater than -1. This has the effect of drastically increasing the average age of GHZ stars in our study and suggests that if we eventually encounter other intelligent civilisations, they are likely to be far more advanced than previously thought.

## § 4.2 Limitations

### § 4.2.1 Timescales for the Evolution of Life in the Galaxy

The timescales for the evolution of life on other planets are not very well constrained as we currently do not know of any life outside of Earth. Although we know roughly when this occurred on Earth we do not know whether this is representative of the galactic mean. There are many uncertainties because of factors such as: extinction events (which may help or hinder the development of life), the location of a potentially habitable planet in the HZ and the mechanisms and timescales which govern the oxygenation of exoplanetary atmospheres. All these things suggest that the timescales for the evolution of life on other planets could vary considerably. When considering intelligent life these timescales are even more uncertain because we are still uncertain about what sequence of events led to our own evolution and there are likely to be a wide variety of alternative evolutionary pathways for intelligence to develop.

In the face of all these unknown factors the most practical approach was to assume that the timescales for intelligent life to evolve here on Earth is representative of the mean for the entire Galaxy. Even though this is unlikely to be the case it would have been unjustified to assume any different based upon the little information we have about the evolution of life in the Galaxy at the present time. As we continue to explore space we may eventually discover some form of basic life on Mars or the moons of the planets in the outer Solar System which could tell us more about how life originated in our Solar System and therefore allow us to constrain these parameters further.

## § 4.2.2 'Missing Physics' in our Galactic Chemical Evolution Model

There are a number of properties and processes of the Galaxy which we were unable to model in this study because of time constraints and the fact that it would have severely increased the amount of computing power required. If these 'missing physics' had been incorporated into our model it is likely that our model would have better reproduced the properties of the Galaxy estimated in other studies. These 'missing physics' include: the infall rate of gas from extragalactic space, the Galaxy's spiral structure, the vertical component of the Galaxy, the motions of stars and radial gas inflows.

The infall rate is of particular importance because including this in our model would mean we could reduce the initial mass of gas which would reduce the SFR and therefore the SNe rate early in the Galaxy's history. This would reduce the metallicity of the Galaxy (especially in the first few gigayears of the Galaxy's formation) and possibly increase the fraction of unsterilised star systems at the Sun's radial position, therefore producing more realistic quantities than our model currently yields. However, it may not have a significant effect on the overall habitability of the Galaxy because although less stars would likely be sterilised, more stars would be deemed uninhabitable because of their lower metallicity. Therefore, it is likely that including infall rate in our model would have the effect of shifting our GHZ closer to the centre of the Galaxy making it more in alignment with Lineweaver et al. (2004) GHZ but without significantly altering the predicted number of PHSS. The inclusion of radial gas inflows on the other hand is likely to increase the total number of PHSS in our GHZ based on the results of Spitoni et al. (2014) who found that including radial gas inflows in their GHZ model increased the maximum number of stars hosting habitable planets in the Milky Way by 38% compared with the "classical" model which did not include radial gas inflows.

Another important property of the Galaxy is its spiral structure. In our model we account for the exponential decline in the mass of gas and subsequently the mass of stars as we move farther from the galactic centre, using equation (1). However, accompanying this exponential decline with radius is the fact that stars are also preferentially located within the spiral arms which stretch out from the galactic centre. This means that the highest density regions of stars and gas at each radial distance from the galactic centre are located at the centre of the spiral arms, with the density decreasing the farther from the spiral arms you move. Consequently, we would expect the SNe rate and metallicity to increase with proximity to the spiral arms suggesting that if a star is born too close to a spiral arm it may not be habitable due to too frequent SNe sterilisations and if a star is born too far from a spiral arm it may not be habitable due to its metallicity being too low for Earth-sized planet formation. We therefore hypothesise that habitability trends with respect to distance from each spiral arm are likely to be similar as habitability trends with respect to distance from the galactic centre. The Sun is located between two spiral arms in a region of lower stellar density, suggesting that if we had modelled the Galaxy's spiral structure the predicted fraction of unsterilised star systems born at the same time and at the same radial position as the Sun may have been higher than the 4% estimated here.

As well as all these additional galactic properties and processes, which if included in future studies would improve the accuracy of our results, there are also some stellar factors which we did not include in the current study such as: the inclusion of other stellar types which may be able to support complex life (F, K and M-type stars (Gowanlock et al. 2011)) and other high-energy events which could cause extinctions in the Galaxy such as black holes, gamma ray bursts (GRB) and Neutron stars. Prantzos (2008) notes that GRB are more powerful than SNe but also much rarer in frequency. This statement is supported by Scalo and Wheeler (2002) who estimated that there are more than 30,000 SNII per GRB with the rate of GRB causing a biologically significant event being ~100-500 $Gyr^{-1}$ in our Galaxy, much lower than the predicted SNe rates. Therefore, it is unlikely that these events contribute significantly to the fraction of stars sterilised in the Galaxy and their effects can be said to be negligible in comparison with SNe. More research is needed to determine the possible dangers that neutron stars and black holes could pose to exoplanetary biospheres, however, these events are also much rarer in frequency compared to SNe, therefore we would not expect them to contribute significantly to the fraction of star systems sterilised in the Galaxy. The inclusion of other stellar types in this study would almost certainly increase the total number of PHSS predicted by our model. However, conditions on planets orbiting other types of stars are likely to be much different compared to planets orbiting G-type stars. This means that any life which arises on these planets is likely to evolve under very different timescales than what has occurred here, therefore modelling the evolution of life on planets surrounding these stars will be much more uncertain. Nevertheless, including other stellar types in future habitability studies would be an interesting endeavour and once we know more about the prevalence of Earth-sized exoplanets around these stellar types they should be included in future habitability studies to provide an upper bound on the number of potentially habitable planets in the Galaxy.

It is hard to predict whether the inclusion of all the 'missing physics' would significantly increase or decrease the size of the GHZ and the average ages of the PHSS. This is because the habitability of the Galaxy is a multifaceted problem dependent on many different factors, therefore it is very difficult to predict what the net effect of including all of these factors would be. What is clear, however, is that there are a lot of ways that our model could be improved - providing much motivation for future research. The field of Astrobiology is still in its infancy, as observational techniques improve and more research is done in relation to exoplanets and galactic habitability, the factors described in this paper will be better constrained and we will be able to draw more accurate conclusions about life in our Galaxy.

## § 4.3 Implications

Our study has demonstrated that there may be as many as $6.2 \times 10^6$ PHSS in the Milky Way. Using the recent results of Petigura et al. (2013), this suggests that there may be as many as ~1.4 million Earth-sized HZ planets in our Galaxy alone (22% of $6.2 \times 10^6$). Not all of these planets will be suitable for the evolution of complex and subsequently intelligent life because of current uncertainties regarding the fraction of these HZ planets which may have suitable atmospheres and geophysics. Nonetheless, this implies that there are many potential homes for intelligent life to have evolved and with the next generation of exoplanet searching telescopes planned for launch over the next couple of decades these uncertainties will eventually be addressed (JWST, n.d., Doyle, 2014, Young, 2015). Furthermore, we have estimated that ~77% of PHSS are older than our Sun with their average age being ~7.68 Gyr, ~3.13 Gyr older than our Sun. This implies that if other intelligent civilisations have arisen in our Galaxy, the majority of them would have had a much longer time to evolve than we have, irrespective of how rare or common their existence may be.

## § 4.3.1 The Fermi Paradox

A key issue when it comes to discussing the possibility of intelligent life in our Galaxy is that of the Fermi Paradox - if there are so many extraterrestrial civilisations (ETC), as we would expect, then "where are they?" (Fermi, 2007). Proposed explanations for the Fermi Paradox normally fall into two categories: explanations as to why ETC are very rare or non-existent and explanations as to why existing ETC have not come here. Explanations falling into the first category seem unlikely given the findings of this study and others which predict that there may be many millions of planets able to support some sort of biological life in our Galaxy alone (~0.06% of stars (the current study), ~0.3% of stars (Gowanlock et al. 2011), ~10% of stars (Lineweaver et al. 2004)). With so many potential homes for life we would expect intelligent life to have evolved in at least some cases. However, with the many uncertainties surrounding the evolution of intelligent life these explanations cannot be entirely ruled out.

Explanations falling into the second category are often regarding the difficulties of interstellar travel. Given the findings of this and Lineweaver et al. (2004) study that the average habitable planet is predicted to be billions of years older than the Earth, we would expect any ETC older than us (and therefore likely to be engaging in space exploration) to be on average billions of years more advanced than we currently are, it therefore seems extremely unjustified to place restrictions on their level of technological ability. This is because their technologies, societies, moral values etc. would likely be unimaginable and possibly even incomprehensible to a human mind. In a remotely similar scenario, how well would an early hominid (living ~8 million years ago) be able to envision human society today? Furthermore, a number of potential ways to travel faster than the speed of light have already been proposed, based upon our current understanding of the laws of physics (for a good overview of these potential technologies see the discussion by theoretical physicist Michio Kaku (Kaku, n.d.)). With NASA and others already starting to look into these technologies (Marcus, 2009, White, 2011, Hambling, 2014), it is perfectly reasonable to think that we may in fact be able to travel to the closest stars even within the next 1000 years (and almost certainly within the next million years). We would therefore expect ETC, billions of years more advanced than we currently are, to have engaged in interstellar travel and started exploring and colonising the Galaxy a very long time ago (probably before animal life had even arisen here on Earth, ~750 million years ago (Gowanlock et al. 2011)).

These conclusions, based on the predictions of this study, suggest that alternative explanations to the Fermi Paradox should be sought. Explanations which allow for the existence of ETC with the technological ability to engage in interstellar travel (and therefore come to our Solar System if they wanted to) but do not choose to reveal their presence to humanity at this time for one reason or another. Consequently, the Fermi Paradox has now become profoundly more difficult to answer as it now involves predicting the motives of advanced ETC, a highly challenging feat.

## § 4.3.2 Implications for Studying the Habitability of Other Galaxies

When testing our model for different values of $r_d$, $\Sigma_0$ and A we gained an insight into how habitability may change between galaxies with different galactic structures. Generally we found that increasing A had the effect of increasing the surface density of stars, decreasing the surface density of gas and decreasing the SFR surface density, whilst the number of SNe sterilisations and metallicity remained largely unaffected. This was as expected because A controls how much gas is converted into stars via equation (2). Therefore, external galaxies found to be following a Schmidt-Kennicutt law with larger values of A would be expected to have a higher surface density of stars and therefore more PHSS but located within a similar sized GHZ.

Increasing $\Sigma_0$ had the effect of increasing the gas mass, stellar mass and SFR at all times and positions in the Galaxy as well as increasing the number of SNe sterilisations and metallicity. This is because $\Sigma_0$=sets the initial central surface density of gas so if there is initially more gas near the Galactic centre more gas will be converted to stars, which explode as SNe therefore increasing the overall metallicity. Consequently, we should expect galaxies with a higher central surface density to be less habitable towards their galactic centre due to an increased chance of SNe sterilisations but possibly more habitable towards the outskirts of the galaxy where the stellar density is lower, as a higher metallicity means more Earth-sized planets would form.

Increasing $r_d$ meant more of the stars and gas in the Galaxy was concentrated towards the galactic centre, resulting in a larger number of SNe sterilisations (especially towards the galactic centre because of the increased surface density there) and reducing the overall number of PHSS. For this reason we would expect galaxies with larger scale lengths to be less habitable than galaxies with smaller scale lengths. This is because our study indicates that SNe sterilisations are the dominant factor in determining a GHZ and the fraction of PHSS in a Galaxy and the fraction of SNe sterilisations is directly dependent on the surface density of stars, which is correlated with the scale length. It is therefore justified to conclude that the fraction of PHSS is inversely proportional to the scale length of a galaxy.

## § 4.4 Concluding Remarks

We have successfully predicted where and when PHSS have formed over the lifetime of our Galaxy allowing us to estimate the age distribution of potential intelligent life in the Milky Way. We have estimated that a GHZ first emerged ~11 billion years ago which has widened with time as metallicity has increased in the outer Galaxy and the SNe rate has decreased and is now centred at ~10 kpc. Our GHZ is located farther from the galactic centre than previous studies (Lineweaver et al. 2004, Gowanlock et al. 2011) have predicted, suggesting that the region of our Galaxy most likely to support the evolution of complex life is located farther from the galactic centre than previously thought. Furthermore, the majority of PHSS are much older than our Sun suggesting that if we eventually discover ETC in our Galaxy they would most likely be much more advanced than we currently are and located more towards the outskirts of the Galaxy. Specifically we predict that ~77% of PHSS are on average ~3.13 Gyr older than the Sun. This means that the majority of intelligent life in our Galaxy is likely to be billions of years more advanced than we are, implying that we are one of the younger civilisations in the Milky Way. It is difficult to draw conclusions about the nature of ETC from this because a billion years is such a vast length of time. For instance modern humans are predicted to have evolved ~100,000 years ago and the dinosaurs are thought to have gone extinct ~65 million years ago which means that we have no evolutionary timescales of this length for comparison. Nevertheless, it is likely that ETC would be immensely different from our own civilisation and their technologies would probably appear like magic to us.

There are a number of improvements which could be made to our model such as including all of the 'missing physics' discussed in section 4.2.3 and correctly modelling SNIa. However, it is unclear what the net effect of including these variables would be on our results. Therefore, future research is needed to more accurately define the GHZ and age distribution of intelligent life, which would improve the predictions made here. We have demonstrated that PHSS have emerged over a wide range of galactic radii and times. Overall we estimate that ~0.06% of G-type stars may harbour an Earth-sized HZ planet, corresponding to ~1.4 million potential homes for life. It would be interesting to see how the inclusion of F, K and M-type stars would affect the results of this study. Although the evolution of life around these stars is likely to be quite different than what has occurred here on Earth, it is important that we attempt to include them in future galactic habitability studies as this would provide an upper bound on the total number of PHSS in our Galaxy.

Taking all this into consideration it could be concluded that ETC most likely exist in our Galaxy and because of their predicted age in comparison to us are likely to be immensely more technologically advanced than we currently are. Therefore, they would almost certainly have the ability to engage in interstellar travel if they chose to do so and it is likely that at the present time the entire Galaxy has been explored and possibly even colonised (considering the results of Kuiper and Morris (1977) who estimated that advanced ETC would be able to completely colonise the Galaxy in a relatively short period of time (~5 Myr)). We should not be disheartened by the lack of incontrovertible evidence of ETC, bearing in mind that absence of evidence is not evidence of absence, which is often overlooked in regards to the Fermi Paradox. Therefore, the possibilities that we are being ignored, avoided or discreetly watched are very plausible (Kuiper and Morris, 1977). Given that these ETC are likely to be on the order of billions of years more advanced than we are it seems unjustified to speculate about why this may be. Instead it could be concluded that they most likely do not want their presence to be known to us yet, possibly for reasons beyond our current level of understanding.

# **Appendices**

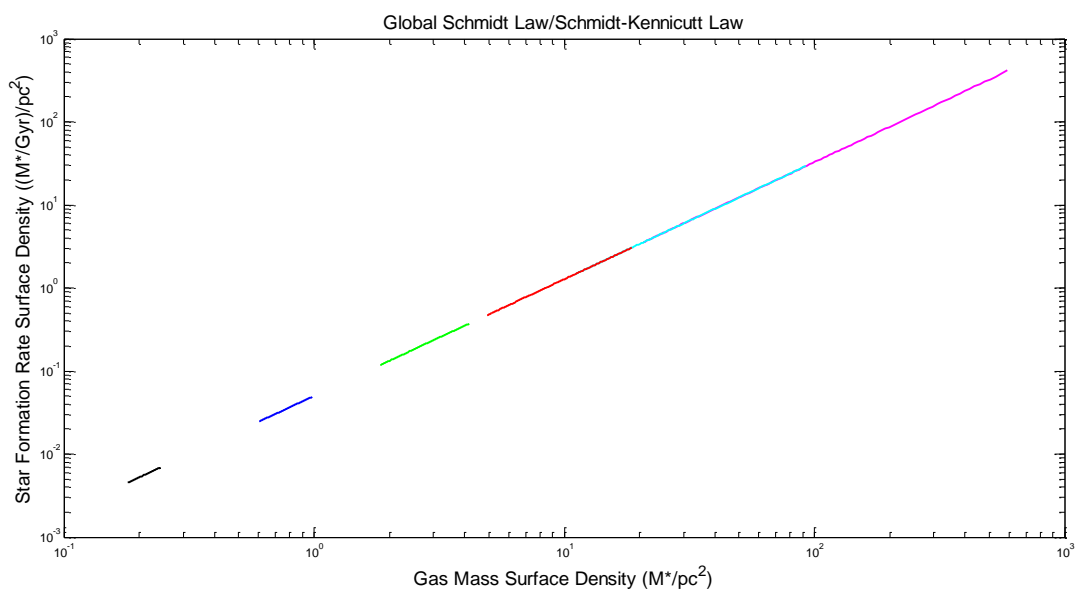

Appendix A: A plot demonstrating how our model follows the Schmidt-Kennicutt law.

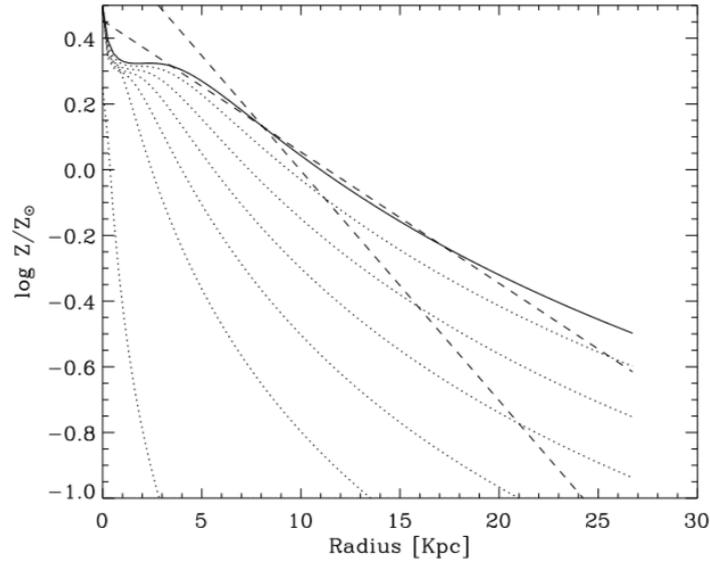

**Figure 11.** Radial metallicity distribution of the gas after 2, 4, 6, 8, 10, 12 and 13.6 $Gyrs$ (thick solid line). The two dotted straight lines indicate a slope of $-0.07 dex kpc^{-1}$ and $-0.04 dex kpc^{-1}$. The metallicity gradient has been steeper in the past. The present day metallicity gradient of the stars is shown by the thick dot-dashed line.

Appendix B: the metallicity profile found by Naab and Ostriker (2006) model of the Milky Way, for comparison with figure 2.

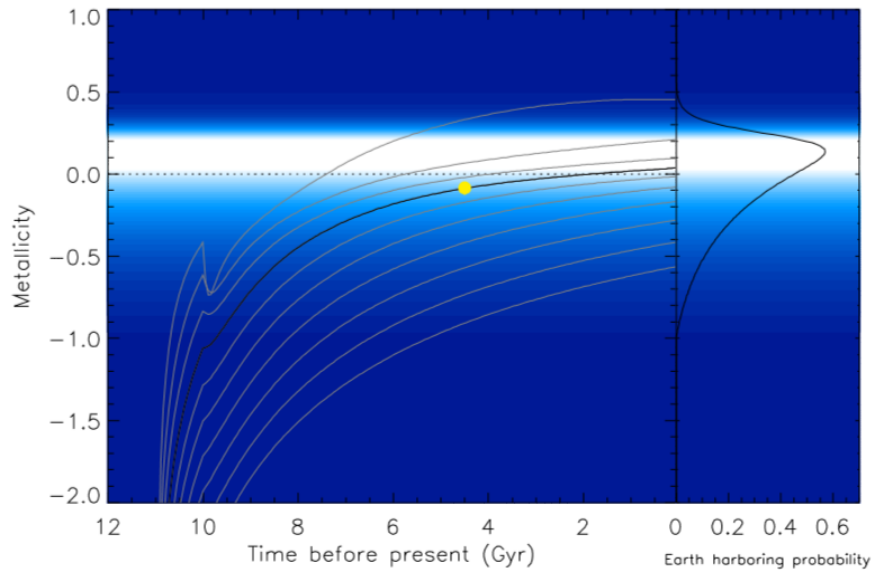

**Fig. 1.** The buildup of metals in our Galaxy as a function of time predicted by our simulations. Metallicities at different Galactocentric distances can be compared with the probability of harboring terrestrial planets as a function of the metallicity of the host star [**right**, see (7) for details]. Galactocentric distances from 2.5 kpc (upper curve) to 20.5 kpc (lower curve) are shown in 2 kpc increments. The yellow dot indicates the Sun's time of formation and Galactocentric distance of 8.5 kpc. The inner Galaxy accrues metals early and rapidly because of a high rate of star formation, whereas the most distant regions remain deficient in the metals needed to form terrestrial planets. The metallicity is the log of the ratio of the amount of iron to hydrogen in the stars relative to the Sun.

Appendix C: Lineweaver et al. (2004) metallicity profile, for comparison with figure 3.